\documentclass{PoS}
\usepackage{graphicx}

\title{Optical timing studies of isolated neutron stars: Current Status}

\ShortTitle{Neutron star timing}

\author{\speaker{Roberto P. Mignani}\thanks{The author acknowledges OPTICON  for financial support for the conference participation and thanks the ESO Visitor Office in Santiago for hospitality where part of this paper was finalised.} \\
        Mullard Space Science Laboratory - University College London \\
        E-mail: \email{rm2@mssl.ucl.ac.uk}}

\abstract{Being fast rotating objects, Isolated Neutron Stars (INSs) are natural targets for high-time resolution observations across the whole electromagnetic spectrum.  With the number of objects detected at optical (plus ultraviolet and infrared) wavelengths now increased to 24, high-time resolution observations of INSs at these wavelengths are becoming more and more important.  While classical rotation-powered radio pulsars, like the Crab and Vela pulsars, have been the first INSs studied at high-time resolution in the optical domain, observations performed in the last two decades have unveiled potential targets in other  types of INSs which are not rotation powered, although their periodic variability is still related to the neutron star rotation.  In this paper I review the current status of  high-time resolution observations of INSs in the optical domain for different classes of objects: rotation-powered pulsars, magnetars, thermally emitting neutron stars, and rapid radio transients, I describe their timing properties, and I outline the scientific potentials of their optical timing studies.  }

\FullConference{High Time Resolution Astrophysics IV - The Era of Extremely Large Telescopes - HTRA-IV,\\
		May 5-7, 2010\\
		Agios Nikolaos, Crete, Greece}

\begin{document}

\section{Introduction}

Isolated Neutron Stars (INSs) are typically identified through the characteristic signature of pulsed electromagnetic radiation, modulated at their spin period.  Being fast-rotating objects, with periods between a few milliseconds up to a few seconds, INSs come out as atural targets for high-time resolution observations across the whole electromagnetic spectrum.  

Pulsations from INSs are mostly detected in the radio band, where  $\sim 1800$ of them have been now discovered as radio pulsars.  However, a growing fraction of INSs is also observed to pulsate at higher energies, with $\sim 80$ and $\sim 60$ (and counting) INSs detected as X-ray and/or  $\gamma$-ray pulsars, respectively  (Becker 2009; Abdo et al. 2010). These include rotation-powered INSs, which can be both radio loud, like the radio pulsars and the peculiar Rotating Radio Transients (RRATs), and radio-silent, like the prototype Geminga, and other classes of INSs, some radio loud, some radio silent, like the magnetars, the X-ray Dim INSs (XDINSs), the Central Compact Objects (CCOs) in supernova remnants (SNRs),  which are not powered by rotation but by other mechanisms yet poorly understood, although their periodic variability is obviously still related to the neutron star rotation.  So far, $\sim 60$ radio-silent INSs have been discovered through X and $\gamma$-ray observations.  
With the number of INSs of all types detected in the optical,  ultraviolet (UV) and infrared (IR) domain (UVOIR for short) now increased to 24 (see Mignani 2010 for an updated review), high-time resolution observations at these wavelengths are becoming more and more important.  While rotation-powered radio pulsars, like the bright Crab ($V\sim 16.5$) and Vela ($V \sim 23.6$) pulsars and PSR\, B0540$-$69 ($V \sim 22.5$) in the Large Magellanic Cloud (LMC), have been the first INSs studied in the optical at high-time resolution, the advent of new observing facilities,  the {\em HST} and the 8m-class telescopes like the {\em VLT}, and of new instrument/detectors, like e.g. {\em ULTRACAM} (Dhillon et al. 2007), {\em OPTIMA} (Kanbach et al. 2008), { \em GASP} (Collins et al. 2009), and {\em  Iqueye} (Naletto et al. 2009), have allowed  to search for, and study, UVOIR pulsations from several objects, mostly rotation-powered pulsars and magnetars.  

This review is organised as follows. In Section 2 I will  present a short description of the different INS types discovered so far. In Section 3 I will briefly describe the case for optical timing studies of INSs and discuss related observational issues.  I will then summarise the state of the art of  timing studies of different classes of isolated neutron stars with UVOIR counterparts: rotation-powered pulsars (Section 4), magnetars (Section 5), XDINSs (Section 6). For each of them, I will describe their timing properties, discuss the physical processes behind the observed pulsations and the scientific potentials of optical timing for the understanding of neutron star physics. In Section 7, I will briefly discuss current perspectives for INSs with no secured UVOIR counterparts: RRATs and CCOs. In Section 8, I will outline possible short/mid-term scientific goals in UVOIR timing studies of INSs which can be realistically pursued with the current telescopes/instruments, like the {\em HST} and the {\em VLT}, while waiting for the next generation of extremely large telescopes, like the European 42m {\em E-ELT}.

\section{The isolated neutron star bestiary}

Interestingly enough, the discovery of pulsed high-energy emission has allowed to unveil the existence of different types of INSs, most of  which, at variance with radio pulsars,  are radio-silent. Moreover, some of them are not powered by the neutron star rotation, according to the classical magnetic dipole model (Pacini 1968; Gold 1968), but by mechanisms not yet completely understood and feature a complex phenomenology.  So far, five (possibly six) different classes  of INSs have been recognised through the detection of their pulsed multi-wavelength emission, including both radio-loud and radio-silent objects.

The first, and more numerous class ($\sim 1800$ objects), is that of the radio pulsars (Lorimer 2009). They span a large range of rotation periods, from a few milliseconds up to a few seconds, age (from a few kyears up to a few Gyears), magnetic fields (from $10^{9}$ to $10^{13}$ G), and rotational energy loss rate (up to $\sim 10^{38}$ erg s$^{-1}$)\footnote{According to the magnetic dipole model, the pulsar age $\tau$ and magnetic field $B$, as well as the rotational energy loss rate $\dot{E}$ are inferred from the value of the period $P$ an its period derivative $\dot{P}$, where $\tau \sim P/2\dot{P}$, $B \propto \sqrt(P ~ \dot{P})$, and $\dot{E} \propto P/ \dot{P}^2$ under the assumption that the neutron star spin period at birth was much shorter than the current period.}.  Radio pulsars are apparently powered by the neutron star rotation,  hence generally referred as rotation-powered pulsars (RPPs). Several of them are also detected to pulsate at high energies (X and $\gamma$-rays), showing characteristic power-law (PL) spectra produced by the emission of synchrotron/curvature radiation from relativistic particles in the neutron star magnetosphere, although pulsed emission from hot polar caps, with a blackbody (BB) spectrum, is also observed in a few  radio pulsars.   In both cases, the emitted radiation is  beamed around the magnetic axis of the neutron star.  Radio pulsars are characterised by a steady, i.e. not bursting and not transient, multi-wavelength emission.  A particular case of radio pulsars is represented by the so-called High-B Radio Pulsars (HBRPs), a handful of objects (5 so far) whose magnetic field is higher than those of most radio pulsars and larger than the quantum critical value ($B_C \sim 4.33 \times 10^{33}$ G), above which radio emission is expected to be suppressed according to standard theories. These high fields make them similar to another class of INSs, the so-called magnetars (see below), although they are markedly different in their emission properties. Interestingly enough, at variance with other few kyears old radio-pulsars they feature X-ray pulsations from hot polar caps, while magnetospheric high-energy emission has been only recently detected in $\gamma$-rays by the {\em Fermi} Gamma-ray Observatory.  To the group of rotation-powered pulsars belong also a sub-group of  INSs which are radio silent, like Geminga, but which are similar in both their high energy emission and in their spin down parameters to radio loud pulsars. Many more radio silent rotation-powered pulsars  are now being discovered in $\gamma$-rays by observations performed with {\em Fermi}  (e.g., Abdo et al. 2009). 

A second class of pulsating INSs is that of the magnetars. These are typically radio-silent INSs with  longer pulsation periods (2-11 s) with respect to radio pulsars,  showing bursting and transient X-ray and/or soft $\gamma$-ray emission (see Mereghetti 2008 for a review). While $\gtrsim 1800$ rotation-powered pulsars have been discovered from multi-wavelength observations, only $\sim 15$ magnetars are known so far.  Being mostly radio-silent, they escape detection from radio surveys while, at variance with  radio-silent Geminga-like INSs, their location makes it difficult to resolve their high-energy $\gamma$-ray emission (if any; Rea et al. in preparation) from the diffuse galactic plane emission, where most of them reside. In the X-rays, it is difficult to disentangle them out from the $\sim$ 500\,000 of sources which populate the X-ray sky, also due to their relatively low quiescent luminosity. Magnetars are thus singled out mainly through their transient and flaring X-ray/soft $\gamma$-ray emission following up on which, pulsations are discovered. Magnetars are also persistent X-ray sources with luminosities by far too high to be powered by the neutron star rotational energy  loss $\dot{E} \approx 10^{32}-10^{33}$ erg s$^{-1}$.   These INSs, of which two possibly different flavours exist, i.e. the Soft Gamma-ray Repeaters (SGRs) and the Anomalous X-ray Pulsars (AXPs), would be instead powered by their hyper-strong magnetic fields of $10^{14}-10^{15}$ G, where field twisting, following crustal fractures on the neutron star surface, and decay would explain the magnetar bursting and persistent emission, respectively.  Their  spectra are characterised by a combination of a BB from hot polar caps and a PL extending towards the hard X-ray/soft $\gamma$-ray energy range, which are responsible for the pulsed high-energy emission. Recently, two magnetars have shown transient  radio emission, apparently not correlated with the X-ray one.  Magnetars are typically young neutron stars, with spin-down ages of a few kyears only, confirmed, in a few cases, by the association with young SNRs. 

A third class (7 objects) is that of the X-ray Dim INSs (XDINSs) which earned their name from their relative faintness in the ROSAT All Sky Survey data (see Haberl et al. 2007 for a review).  XDINSs are radio-silent, although for some of them detections at low radio frequencies have been claimed,  and they are detected only through their purely thermal and steady BB X-ray emission from the cooling neutron star surface. Most of them feature long period ($P=3-11$s) and shallow X-ray pulsations presumably produced by hot and large polar caps. They are probably a few Myear old, as suggested by the lack of associated SNRs and, in those cases where a $\dot{P}$ is measured, by their spin-down age.  Some of them are possibly endowed with high magnetic fields ($B \sim 10^{13}-10^{14}$ G), inferred from the detection of spectral features in the X-ray spectrum and, in some cases,  also from  the spin-down parameters.  On the other hand, they appear to be characterised by a  very low $\dot{E} \approx 10^{30}$ erg s$^{-1}$ which makes it difficult for the neutron star rotation to power detectable  multi-wavelength emission.

One of the more puzzling classes of INSs, whose nature has been possibly understood only recently, is that of the Central Compact Objects (CCOs) in SNRs. CCOs ($\sim 10$ objects) are radio silent and only observed in X-rays, where pulsations have been detected  for 3 of them   (see De Luca et al. 2008 for a review). They are supposedly young INSs, being associated with a few kyear old SNRs, although for the very few cases for which the $\dot{P}$ has been measured, or constrained, the spin-down age turns out to be several orders of magnitude larger than the SNR age.  This suggests a scenario where CCOs are INSs born rotating close to their current period  and with very low magnetic fields ($B<10^{11}$ G) and rotational energy loss rate ($\dot{E} < 10^{32}$ erg s$^{-1}$). Their X-ray emission is indeed thermal and produced by hot polar caps, possibly re-heated by accretion from a debris disk formed out of the supernova explosion, allowed by the low neutron star magnetic field.

The last discovered INS class is that of the Rotating RAdio Transient (RRATs). They are transient, long period ($\sim 0.4-7$ s), radio pulsars which feature extremely short ($<50$ ms) radio outbursts, when  they outshine in brightness all other radio pulsars, after which they turn invisible for up to a few hours (see Mc Laughlin et al. 2009 and  Keane, these proceedings, for reviews). This bursting behaviour makes them more similar to the magnetars than to, e.g. the restricted class of intermittent radio pulsars (Lyne 2009), although they do not necessarily resemble magnetars in their spin-down parameters. Only a tiny minority of the known RRAT population ($\sim 20$ objects) has been detected at wavelengths other than radio. In particular, only a handful of objects have been detected in X-rays, of which only one has been found to pulsate. Their age varies considerably, suggesting that their properties are not related to a peculiar phase of the neutron star evolutionary pattern, although it has been suggested that some RRATs might be old  radio pulsar undergoing convulse emission activity before turning radio silent. 
 
The existence of different types of INSs has still to be explained in a coherent neutron star fomation/evolution scenario and a possible link between different types has to be  investigated (see, e.g.  Kaspi 2010), although their relative distribution in a $P-\dot{P}$ diagram (Fig.\ 1) suggests possible evolutionary paths, eventually linking magnetars, RRATs, and XDINSs. In particular, it is not clear whether the characteristics of the progenitor star,  the supernova dynamics, or processes occurring in the post-supernova phase, establish the ultimate fate of a massive star, driving a newborn neutron star to become a radio pulsar, a magnetar, or whatever different object it can turn into.   

\begin{figure}
\centering
\includegraphics[width=9.0cm,clip]{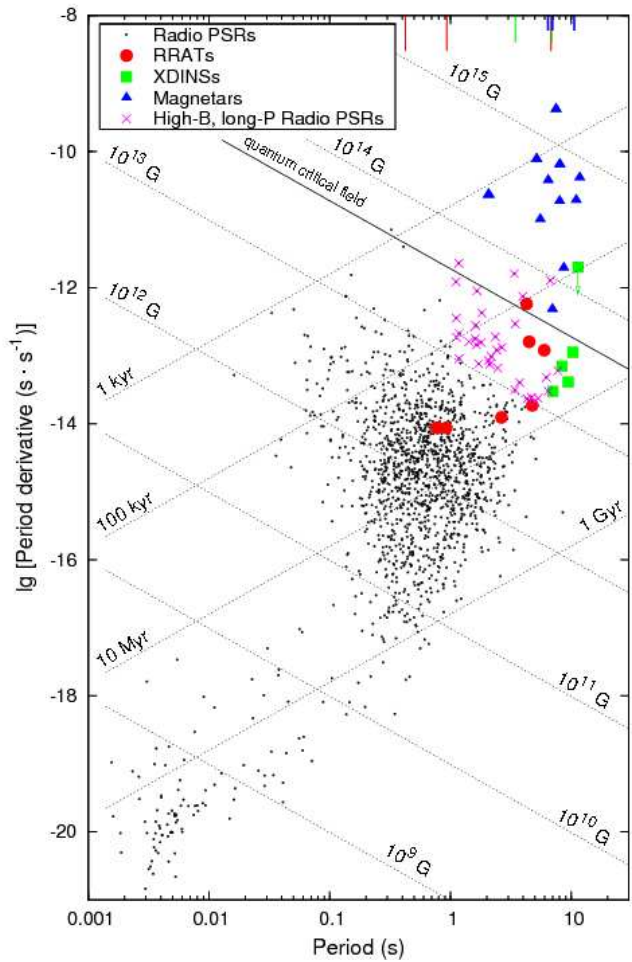}
\caption{$P-\dot{P}$ diagram for all known INSs (see legenda). }
\label{p-pdot}       
\end{figure}

\section{Optical timing of isolated neutron stars: expectations and observations}
  
All the above INS types are also obvious targets for timing observations at UVOIR wavelengths.  Different emission mechanisms can be responsible for the production of UVOIR pulsations.  Synchrotron or other non-thermal processes in the neutron star magnetosphere can produce UVOIR emission near the magnetic poles, which would  be highly collimated and, thus, pulsed.  These mechanisms are expected to be at work in INSs which are characterised by strong magnetospheric activity at other wavelengths like, e.g. rotation-powered pulsars, magnetars, and RRATs.  In most cases, the UVOIR light curves would be expected to feature sharp, possibly double peaked, profiles reflecting the higher density of the magnetic field lines close the magnetic poles and the magnetic pole viewing geometry. Phase shift with respect to, e.g. the  X-ray and $\gamma$-ray light curves would be expected if the UVOIR emission comes from different regions of the neutron star  magnetosphere. Like in the X-rays, optical/UV pulsations could also  be produced, in principle, by thermal emission from the neutron star surface provided, e.g. that its temperature distribution is not isotropic.  In this case, optical/UV pulsations might originate either from the hot polar caps producing the X-ray pulsations, with the optical spectrum being the Rayleigh-Jeans (R-J) tail of the X-ray BB, or from other regions which are colder than the polar caps but still hotter than the bulk of the neutron star surface.  In this case,  the optical/UV light curves would feature more or less shallow profiles depending on the actual size of the emission region. Phase shifts with respect to the X-ray light curve would also be expected  if the optical/UV and the  X-ray emission come from different regions on the neutron star surface.  These types of optical/UV pulsations might thus  be produced from thermal X-ray emitters, like the XDINSs and the CCOs.   UVOIR pulsations could also be indirectly produced from the reprocessing of pulsed X-rays in a debris disk surrounding the neutron star, regardless of the nature (magnetospheric/thermal) of the X-ray pulsations and on the INS type.  In this case, wider profiles of the UVOIR light curves would be expected with respect to the X-ray one due to the radiation smearing in the disk, as well as phase lags due to the neutron star-disk travel time. Thus, timing observations at UVOIR wavelengths, including phase-resolve polarimetry and spectroscopy, provide important element for the understanding of their global emission properties, for the study of the neutron star magnetosphere and of the neutron star cooling surface, and for investigating the presence of debris disks. 

UVOIR timing observations obviously come as follow-ups of  INS detections obtained through  imaging observations, with the only exception being represented by PSR\, B0540$-$69 in the LMC, for which pulsations from a yet unidentified counterpart were detected first (Middleditch \& Pennypacker 1985). Up to now, 24 INSs have been identified at UVOIR wavelengths, albeit with different degrees of confidence. The list includes 12 rotation-powered pulsars, 5 XDINSs, and 7 magnetars (see, e.g. Mignani 2010 for an updated review). Moreover,  candidate counterparts have been found for the CCO in the Vela Jr. SNR (Mignani et al. 2007a), for two more magnetars (Testa et al. 2008), and for  RRAT\, J1819$-$1458 (Rea et al. 2010).  However, out of the 24 INSs with an UVOIR identification only 8, i.e. exactly one third of the sample, have been found to pulsate at these wavelengths, of which 5 are rotation-powered pulsars and 3 are magnetars.  

Why are UVOIR pulsars so rare?   A first reason is physical. Admittedly, the major limit to UVOIR pulsar detection is represented by the intrinsic object faintness,  with only 3 INSs brighter than $V \sim 25$, which makes them very challenging targets for   timing measurements (see below). Moreover, the Pulsed Fraction (PF) depends on the underlying emission process, being larger in case of magnetospheric emission, resulting in a much sharper and more easily detectable pulsation, and much shallower in case of thermal emission.  However, it is difficult to determine {\it a priori} an expectation value for the PF, which, in turn, makes it difficult to predict the number of counts required to detect pulsations at a given confidence level.  This issue is obviously more important for UVOIR ground-based observations, where the count statistics is strongly affected by the sky conditions.  Last but not least,    depending on the INS spectrum and the interstellar extinction,  pulsations might be observable, e.g. in  the IR but not in the UV, thus making detection chances critically dependent on the observing wavelength and subject to instrument/detector selection effects. A second reason is technical.  In particular, the lower number of photons collected at UVOIR wavelengths makes it impossible for most INSs to perform a periodicity search through a Fast Fourier Transform (FFT). Thus, photons must be folded around a reference period known from measurements obtained beforehand in, radio, X-rays, or $\gamma$-rays. For this reason, the INS must be a stable pulsar, i.e. its period must not be affected by glitches, in which case contemporary ephemeris are needed.      Finally, a third reason is  instrumental.   Being intrinsically very faint objects, UVOIR timing of most INSs  had to wait for the advent of 8m-class telescopes, like the {\em VLT}, although it still suffers, in most cases, for the lack of on-site instruments for high resolution optical timing observations.   On the other hand, the {\em HST} has been equipped with two instruments for high time resolution optical/UV observations: the High Speed Photometer (HSP) and the Space Telescope Imaging Spectrograph ({\em STIS}). However, the {\em HSP} was removed in 1993 during the first servicing mission, after only 3 years of operations,  while the {\em STIS} has been unavailable for 5 years between 2004 and 2009 due to a failure in the on-board electronics. Although in the last decade many instruments for high time resolution UVOIR observations have been built, like, e.g. {\em ULTRACAM} (Dhillon et al. 2007), {\em OPTIMA} (Kanbach et al. 2008), { \em GASP} (Collins et al. 2009), and {\em  Iqueye} (Naletto et al. 2009), most of them are guest instruments managed and operated by the construction consortia and, apart from {\em ULTRACAM}, are not always available to the Community through open time proposals. Moreover, some of them might not be easily portable to all telescopes because of the lack of focii available for guest instruments, or because of  interfacing/operability problems, or other related technical issues.  Telescope management concerns in providing guest instrument support and direct costs for instrument shipping and operation also represent severe limitations.

\section{Rotation-powered pulsars}

The Crab,  identified with the Badee and  MinkowskiÕs "south-preceding star" even before the pulsar discovery,   has been the first rotation-powered pulsars, hence the first INS, to be detected as an optical pulsar (Cocke et al. 1969).  Legend has it, however, that optical pulsations from the Badee and  MinkowskiÕs  star were detected naked eye in the early 1960s by a woman attending an open night at the Flagstaff Observatory, although that claim was never followed up on  (Bell-Burnell 2004). A total of 12  rotation-powered pulsars have been identified so far at UVOIR wavelengths ($V\sim 16.5-28$). In particular, all of them have been detected in the optical, 4 in the near-IR, and 7 in the near-UV.  Most of them are relatively young (ages $<$ 0.5 Myear) and energetic ($\dot{E} > 10^{33}$ erg s$^{-1}$).  Their optical luminosity $L_{opt}$ is found to be proportional to the $\dot{E}$, which means that it is mostly sustained by the neutron star rotation. A PL,  ascribed to synchrotron radiation, is recognisible in the UVOIR spectra of  all rotation-powered pulsars, dominant in the optical/IR, in some cases  coupled with a R-J, ascribed to thermal radiation  from the neutron star surface, dominant in the near-UV.  The relative contribution between the two components apparently depends on the INS age, although no clear turnover is recognisible, nor any evidence of  an optical PL/R-J spectral evolution (Mignani et al. 2007b).  This means that even relatively old pulsars can still be efficiently produce  magnetospheric emission in the optical band. Interestingly enough, breaks in the optical-to-X-ray PL spectra are common  in all rotation-powered pulsars   (Mignani et al. 2010a), suggesting a complex particle density and velocity distribution in the neutron star magnetosphere.  As discussed in the previous section, the presence of a strong PL component in the UVOIR spectra makes pulsations an expected feature of rotation-powered pulsars.  In general, they are the most promising targets for optical timing observations.   Many potential targets are found from pulsar radio surveys, and now from $\gamma$-ray surveys with {\em Fermi} (see Caraveo, these proceedings).   Thus, at a variance with other INS types,  rotation-powered pulsars count on accurate period ephemeris from radio and/or X-ray/$\gamma$-ray observations, which yield the spin-down parameters, including the $\dot{E}$, hence an estimate of the expected optical luminosity. Moreover, radio observations yield the measurement of the distance,  while  X-ray observations provide a measurement of  the interstellar extinction (by fitting the hydrogen column density $N_{\rm H}$ to the X-ray spectra), hence the expected optical brightness.  

So far, pulsations have been detected for 5 out the 12 rotation-powered pulsars with an UVOIR counterpart.  These are the Crab pulsar, (e.g. S\l{}owikowska et al. 2009), the Vela pulsar (e.g., Gouiffes 1998), PSR\, B0540$-$69 (e.g., Gradari et al. 2010),  PSR\, B0656+14 (Shearer et al. 1997; Kern et al. 2003) and Geminga (Shearer et al. 2008).  The first three, which are also the brightest ones,  have all been recently re-observed with {\em Iqueye} (Barbieri et al., these proceedings). All but PSR\, B0540$-$69 are detected as pulsars both in the optical and in the near-UV.   The light curves of rotation-powered pulsars are mostly  double-peaked,  with the two peaks clearly separated in phase by $\sim 0.4-0.6$,  while PSR\, B0540$-$69 features two narrow peaks superimposed on a main, broader, peak. The peaks in the optical light curves  are not always aligned in phase with the radio/X-ray/$\gamma$-ray ones and the optical light curves profiles are often different from those seen at radio and high energies,  suggesting different emission geometries and/or viewing angles.  In particular,  in the case of the Crab pulsar the optical light curve reproduces quite well the $\gamma$-ray and one (also double-peakd), with a phase shift $\Delta \phi << 0.1$ and the two peaks separated by $\sim 0.5$ in phase, while for the Vela pulsar the optical light curve shows broader peaks with respect to the $\gamma$-ray one and is not aligned in phase, having  a much smaller phase peak separation of $\sim 0.2$ in phase.  Differences are also seen between the optical and $\gamma$-ray light curves of PSR\, B0656+14 and Geminga where, in both cases, the $\gamma$-ray peaks lead in phase by $\approx 0.1$ the optical ones. On the other hand, the optical light curve of PSR\, B0540-69 bears a close resemblance with the X-ray one (the pulsar is not detected in $\gamma$-rays) with the first of the two small peaks substantially aligned with the X-ray one.

Optical timing observations of rotation-powered pulsars are crucial to investigate the phenomenon of the giant pulses (GPs), observed for the first time in the radio band (GRPs) in several radio pulsars. Giant pulses are erratic variation  of the peak-to-peak flux ratio which tend to repeat more or less frequently in different pulsars, with the Crab being the more prolific. Interestingly enough, GRPs are typically observed in the pulsars which have the highest value of the magnetic field at the light cylinder radius $r_{LC} = c P/2 \pi$. So far, a similar phenomenon (GOPs) has been observed in the optical only for the Crab (Shearer et al. 2003), thanks to simultaneous observations of GRPs, showing a peak-to-peak variation of $\sim 3\%$ in the optical light curve. The simultaneous occurence of GRPs and GOPs suggests that the optical (incoherent) and radio (coherent) radio emission are linked at some level, at least in the Crab, perhaps due to the phase-alignment of their light curves. Thus, searching for GOPs in other pulsars is extremely important. While the potentially more interesting candidate, PSR\, J0537$-$6910, is still undetected in the optical despite of deep searches with the {\em HST} (Mignani et al. 2005), PSR\, B0540$-$69, the second brightest of the optical pulsars and, of those, that with the highest value of the magnetic field at the light cylinder, is obviously a primary target. 

Pulsar timing in the optical has been employed to infer the value the braking index. This parameter ($n$) is related to the  pulsar spin-down as $\dot{P}=-K~ P^{2-n}$ and, for magnetic dipole emission and for  an aligned rotator  it is  equal to 3.   Optical measurements of the pulsar first an second period derivative $\dot{P}$ and $\ddot{P}$ are crucial for those pulsars which are either radio silent or very faint in the radio band.  This is the case of PSR\, B0540$-$69 in the LMC. The pulsar is the second brightest detected  in the optical ($V\sim 22.5$) and one of the 5 detected optical pulsars (see above).  The pulsar $\dot{P}$ and $\ddot{P}$ have been monitored through the years with different facilities, both from the ground (Manchester \& Peterson 1989; Gouiffes et al. 1992) and from the {\em HST} (Boyd et al. 1995), and several values of the braking index have been obtained. The most recent measurement  has been obtained from  timing observations performed with the {\em Iqueye} instrument mounted at the ESO 3.5m {\em NTT}, yielding $n = 2.047 \pm 0.040$ (Gradari et al. 2010).  This is in good agreement with most published values both at optical an X-ray wavelengths and confirms that the braking index deviates from the canonical value, suggesting that he pulsar  might be subject to an extra torque. 

Timing studies of INSs are not only aimed at the periodicity search but also, if the object brightness allows, at performing phase-resolved spectroscopy and polarimetry. The first rotation-powered pulsar for which phase-resolved spectroscopy has been performed is, obviously, the Crab. Observations performed with the {\em HSP} on {\em HST} (Percival et al. 1993) showed that the light curve profile is wavelength-dependent, with the peak widths and separations larger in the optical with respect to the near-UV.  It is interesting to speculate whether this change of the light curve is related to change of the PL spectrum which occurs between the optical an the near-UV, with a turn-off of the spectral index (Sollerman et al. 2000).  Phase-resolved spectroscopy observations of the Crab performed with the {\em STIS} on {\em HST} (Gull et al. 1998), however, did not find any obvious wavelength dependence of the light curve profile  in the near-UV, somehow, indirectly  strengthening the idea that light curve and spectral changes are related. On the other hand, phase-resolved {\em STIS} observations were also performed for 3 more pulsars: Vela (Romani et al. 2005), PSR\, B0656+14 (Shibanov et al. 2006), and Geminga (Kargaltsev et al. 2005) and showed that the light curve profiles change with the wavelength in the near-UV, allowing to weight the contributions of different emission components, magnetospheric and thermal, to their optical/UV spectra.  

The only rotation-powered pulsar for which phase-resolve polarimetry observations have been repeatedly performed is the Crab pulsar (see S\l{}owikowska et al. 2009 and references therein).  Observations performed with {\em OPTIMA} have clearly shown, confirming and improving on previous results, that both the polarisation degree (PD) and position angle (PA) clearly depends on the pulse phase. In particular,  PD is maximum in the inter-pulse region, and minimum at the peaks, while PA swings before the peaks, with the swing being maximum before the main peak. However, particularly crucial in the case of the Crab pulsar is the subtraction of the DC component. Indeed, {\em HST} images have shown the presence of a bright, non-thermal, emission knot (Sandbergh \& Sollerman 2009) located at $\approx 0.4$ arcsec from the pulsar and unresolved in ground-based images.  The knot can, thus, significantly contribute to the DC polarisation of the Crab. Thanks to their superior spatial resolution, phase-averaged {\em HST} polarisation images of the Crab pulsar have been crucial to measure the knot polarisation which turns out to be $\approx 70 \%$ with respect to the $\approx 5\%$ measured for the pulsar (S\l{}owikowska et al., these proceeings). Indeed,  the total pulsar+knot phase-averaged polarisation accounts for the $\sim 10 \%$ PD obtained by integrating the phase-resolved polarisation values over the pulsar period (S\l{}owikowska et al. 2009).  Phase-resolved polarimetry observations have also been performed for PSR\, B0656+14 (Kern et al. 2003) but, unfortunately, only on $\approx 30\%$ of the period. Nonetheless, these observations showed that PD $\sim$ 100\% in the interpulse possibly suggesting, like in  the case of the Crab, that the polarisation is not maximum at the peaks. 

\section{Magnetars}

The second class of most potential targets for UVOIR timing observations is represented by the magnetars. However, at a variance with rotation-powered pulsars which can be as close as a few hundreds of pc, magnetars are much more distant, with distances typically of few kpc, and most of them are located in the crowded and highly extincted  regions of the galactic plane (up to $A_V \sim 30$), which is a killer for optical/UV observations.  Nonetheless, 6 magnetars have been identified in the optical/IR so far ($K_s \sim 18.5-21.5$), thanks to their brightening following an X-ray burst or flare. 
The nature of the magnetar optical/IR emission is not clear yet. While, e.g. their hard X-rays/soft $\gamma$-ray emission is certainly of magnetospheric origin, very little is known about their optical/IR spectra, whose characterisation is affected by the interstellar extinction and by the object variability. For the magnetars with low measured extinction, however, a PL  yields a good spectral fit to the optical/IR spectrum, suggesting that the optical/IR emission is non-thermal.  However,  the relatively low rotational energy loss rates, $\dot{E} \sim 10^{32}-10^{33}$ erg s$^{-1}$,  would make  rotation-powered synchrotron emission from magnetars intrinsically much fainter than that of, e.g.  radio pulsars.  Interestingly enough, the ratio between the  magnetar IR luminosity and the rotational energy loss rate  $L_{IR}/\dot{E}$, which define the rotation-powered IR emission efficiency,  is $\approx 10^{-3}$, i.e. a factor of $\sim 100,000$ higher than that of rotation-powered pulsars (Mignani et al. 2007c). In particular, this ratio seems to correlate with the magnetic field, which might suggest that the magnetar IR emission is powered, like at high energies, by the magnetic field rather than by the neutron star rotation. An alternative explanation is that this relatively large IR luminosity is due to the contribution of an extra source, like a debris disk around the magnetar. A possible evidence for a debris disk has been found from the {\em Spitzer} mid-IR detection of the AXP 4U\, 0142+61, while for some of the others only upper limits have been obtained (see Mignani 2010 and references therein). 

Magnetar timing observations in the optical/IR, however, provide an important diagnostic. Optical/IR pulsations have been detected, so far, in  only three magnetars.  The first one is the AXP 4U\, 0142+61, found to pulsate in the optical (Kern \& Martin 2002; Dhillon et al. 2005) but not yet in the near-IR (Morii et al. 2009). The optical pulsed fraction $PF_{opt}$ is about 5 times larger than the X-ray one ($PF_X$), which would argue against a disk   origin of the optical emission (see Section 3).  On the other hand,  no significant evidence of a phase lag of the optical pulse ($\Delta \phi = 0.04\pm 0.02$) was found, which would, instead, argue in favour of a disk emission. On the contrary, for the AXP 1E\, 1048$-$5937 (Dhillon et al. 2009) $PF_{opt} \sim 0.7 PF_X$, while there is a marginal evidence of the optical pulse leading the X-ray one ($\Delta \phi = -0.06\pm 0.02$). In both cases, the profile of the optical light curve follows quite well that of the X-ray one in the case of 1E\, 1048$-$5937, with one single peak, while  similarities are less evident for 4U\, 0142+61. All in all, present observations seem to favour a magnetospheric origin at least of the pulsed optical emission, with continuum emission possibly coming from a disk. Conclusions, however, must come with the caveat that comparing non-simultaneous optical/X-ray observations introduce biases due to the observed long-term (and possibly optical) X-ray light curve variability, which depends on the magnetar state.   
Optical/IR pulsations have been detected also  for a third magnetar, SGR\, 0501+4516	 (Dhillon et al. 2010), for which an optical/IR counterpart has been recently discovered (Levan et al. 2010). Interestingly enough, optical pulsations from magnetars have all been observed with {\em ULTRACAM} (see Dhillon, these proceedings). 
Optical pulsations  might have been  detected from a magnetar candidate, the transient X-ray source SWIFT\, J195509.6+261406, observed by {\em OPTIMA} during a very intense optical flaring activity (Stefanescu et al. 2008). In particular, a possible pulsation with a period of 6-8 s has been found  in the strongest flares,  i.e. a value similar to the magnetar periods (see also, Kanbach et al., these proceedings).

\section{X-ray Dim INSs}

After the magnetars, the XDINSs are the INSs which count more UVOIR identifications. So far, 5 XDINSs have been identified in the optical ($V\sim 25.5-28.5$), while none has been detected in the near-IR, despite of several attempts (see Mignani 2010 and references therein). Surprisingly enough, no observation of XDINSs in the near-UV  have been performed so far. While XDINSs are the INS class with the highest score of optical identifications, only thanks to their close distance ($\lesssim 350$ pc) and very low interstellar extinction ($A_V<0.2$), they are very challenging targets for optical timing observations. Indeed, their rotational energy loss rate, $\dot{E} \approx 10^{30}$ erg s$^{-1}$, is too low to power highly-beamed synchrotron optical emission, like in radio pulsars, unless the XDINS optical emission efficiency is at least a factor of $10^3$ higher.  XDINSs are characterised by an optical excess with respect to the extrapolation of the X-ray BB.  In the best studied cases, the optical spectrum follows a clear R-J continuum, which is compatible with thermal emission from a colder fraction of the neutron star surface.  However, determining what fraction of the neutron star surface actually emits the optical R-J relies on accurate distance measurements.  Only in two cases XDINS distances have been obtained through the measurement of the optical parallax, with an uncertainty of $\sim 10-30\%$, suggesting that the optical emission might be produced from the bulk of the neutron star surface. In this case, no optical pulsations would be detectable. Of course, it can not be ruled out that other XDINSs feature  more complex thermal maps, which would make optical pulsations possibly detectable. A possible different case is represented by the XDINS RBS\, 1774, whose optical flux can not be easily explained in terms of thermal emission from the neutron star surface (Zane et al. 2008) but could be related to the neutron star magnetic field, the highest among XDINSs, as it is proposed for the magnetars. In this case, the production of optical pulsations might be more likely.  However, no high time resolution optical observations of XDINSs have been performed so far.

\section{RRATs and CCOs}

For all other INS types no UVOIR counterparts have been detected so far and possible candidates have been found only for very few of them. 
Although a candidate IR counterpart has been detected only in one case (Rea et al. 2010), RRATs are very interesting targets for optical/IR timing observations aimed at searching both for IR outbursts, like the radio ones, and for IR pulsations. Interestingly enough, RRAT\, J1819$-$1458  has been related to magnetars because of its X-ray luminosity, in excess of its rotational energy loss rate, and of its magnetic field, just above the critical quantum limit. In addition, its possible IR counterpart might have an emission efficiency comparable to the magnetar one. Thus, like magnetars, some RRATs might feature a strong magnetospheric IR emission and be detectable as IR pulsars. What matters in the case of RRATs, however, is that, although their integrated IR flux can be quite faint, their IR flux at the peak of an outburst can be orders of magnitude higher, at least if the IR and radio emissions are alike. This would make it more do-able to detect pulsations than in other INSs of comparable integrated brightness, once the IR flux is time-resolved and sampled with a time resolution sufficient to pinpoint each single outburst. So far, a search for IR burst and pulsations has been performed only for RRAT\, J1819$-$1458 (Dhillon et al. 2006), prior to the discovery of its candidate counterpart, but none of them were found. 
For the CCOs, however, the situation is less promising.  Indeed, because of their characteristics, CCOs are very difficult targets for optical/IR timing observations. Their low rotational energy loss rate, $\dot{E} \lesssim 10^{32}$ erg s$^{-1}$, makes their optical/IR magnetospheric emission (if any) intrinsically very faint, while the small size of their hot polar caps would imply that thermal emission would be even fainter. In principle, pulsations from X-ray reprocessing in a debris disk would be a possibility but current upper limits on the disk size and mass (e.g. De Luca et al. 2010) imply a very low disk luminosity. Despite of several observations carried out both with the {\em HST} and the {\em VLT} (see Mignani 2010 an references therein), a candidate IR counterpart has been detected only for the CCO in the Vela Jr. supernova remnant (Mignani et al. 2007a).  However, its faintness ($K_s \sim 21.4$) makes it virtually prohibitive to perform a periodicity search since no reference  period has been measured yet in X-rays, like for most CCOs. 

\section{Summary and current perspectives}

So far, pulsations have been detected from 8 out of the 24 INSs detected at UVOIR wavelengths. They all are either rotation-powered pulsars or magnetars, i.e. all objects which feature magnetospheric emission both in the UVOIR as well as at other energies. Thus, they are the best target for timing observations with the present/future observing facilities.  The suite of instruments currently available for high time resolution observations of INSs in the UVOIR offer a wide range of options, with both ground and space-base telescopes. On the {\em HST}, the refurbished {\em STIS} and the newly installed Cosmic Object Spectrograph ({\em COS}) offer the possibility of performing high-time resolution observations in the near-UV, including phase-resolve spectroscopy. From the ground,  search for pulsations can be performed  with  {\em OPTIMA} (Kanbach et al. 2008), {\em ULTRACAM} (Dhillon et al. 2007), {\em Iqueye} (Naletto et al. 2009), and  {\em GASP} (Collins et al. 2009), although the different time resolution achievable with these instruments make them more suitable for different types of INSs, i.e. either shorter period pulsars or longer period magnetars. Moreover, both  {\em OPTIMA} and {\em GASP} also offer the possibility of performing phase-resolved polarimetry, so far not possible with the other quoted instruments.   Different science goals in high time resolution UVOIR studies of INSs are still to be achieved in the {\em HST} and 8m-class telescope era, exploiting the current generation of instruments.
 
The first goal is, of course, to perform deeper timing observations of already known UVOIR pulsars to better characterise their light curves and to phase-resolve their spectra and polarisation degrees.  
For instance, the dependence of the light curve profile has been studied on the Crab pulsar only.   Multi-band optical light curves can be easily obtained for the Crab pulsar, PSR\, B0540$-$69, and the Vela pulsars with, e.g.  {\em Iqueye} on a middle-size telescope like, e.g. the 3.5 m {\em NTT}, while 8m-class telescopes, like the {\em VLT}, would be required for the fainter PSR\, B0656+14 and Geminga. In particular, a monitoring of the PSR\, B0540$-$69 light curve could make it possible to search for giant optical pulses (GOPs), so far only detected in the Crab pulsar. For the magnetars, multi-band light curve studies can be performed with {\em ULTRACAM}. Moreover, monitoring the optical/IR light curves as a function of the source state would be crucial to investigate disc emission models.  
For both PSR\, B0540$-$69 and the Vela pulsar, phase-averaged PD of $\sim 10\%$ have been measured (Mignani et al. 2007d; Mignani et al. 2010a), but  phase-resolved polarimetry observations have never been performed so far, which are within reach of both {\em OPTIMA} and {\em GASP} at the {\em NTT}.  At the same time, phase-resolved polarisation measurements of PSR\, B0656+14, carried out only partially, and of Geminga, yet to be carried out, are feasible at the {\em VLT}. 
Phase-resolved spectroscopy is currently  possible only with both {\em STIS} and {\em COS} on the {\em HST}, and it is thus strongly biassed  towards the near-UV.  Among the 5 known UVOIR pulsars, the only remaining target is PSR\, B0540$-$69, whose near-UV emission properties are still unexplored (Mignani et al. 2010). 

The second goal is to first search for pulsations from other INSs already detected at UVOIR wavelengths  but not yet as UVOIR pulsars.  Among the rotation-powered pulsars, for instance,  PSR\, B1055$-$52 (Mignani et al. 2010b) is the next best target for near-UV timing observations with either {\em STIS} or {\em COS} on the {\em HST}, together with  PSR\, B1929+10 for which {\em STIS} observations had been tried (Mignani et al. 2002) but with inconclusive results. PSR\, B1509$-$58, the second youngest pulsar with an optical counterpart, is also a prime target for timing observations, to be carried out at longer wavelengths to cope with its much higher extinction ($A_V \sim 5$). Among magnetars, the best candidates for {\em ULTRACAM} observations are probably the AXPs 1E\, 2259+586 and XTE\, J1810$-$197. For the others, unfortunately, the search for pulsations is hampered by the high extinction, which requires high-time resolution instruments optimised for the near-IR, not available yet. 

The third goal, of course, is to follow-up both on already known INSs, once their UVOIR identification is achieved,  and on new INSs discovered through multi-wavelength observations, especially in the high-energy domain. While many INSs can be identified from systematic surveys of unidentified X-ray sources in the large  {\em Chandra} and {\em XMM-Newton} archives, a growing number of INSs is being detected in $\gamma$-rays from the wide-area sky scans performed by {\em Fermi}.  In the radio band, dedicated sky surveys will yield many more RRAT identifications, many of which might still lurk undiscovered.   

These observations will pave the way to future UVOIR timing studies of INSs to be carried out in the era of extremely large telescopes, like  the {\em E-ELT} (Shearer, these proceedings). The {\em E-ELT}  has the potentialities of discovering hundreds of new INSs in the optical/IR, following up on radio and X-ray observations performed with the {\em Square Kilometre Array} ({\em SKA}) and the {\em International X-ray Observatory} ({\em IXO)}, (see Kramer and Becker, these proceedings) as well as on the large database of $\gamma$-ray observations left after the end of the {\em Fermi} mission, enlarging the sample of identified INSs by at least a factor of 5 and filling the gap between the UVOIR and high energy domains. More importantly, the large collecting areas of the {\em E-ELT} and other extremely large telescopes will make it possible to perform high-resolution timing observations as part of the regular UVOIR data analysis, together with spectroscopy and polarimetry, expanding the INS research on the domain most typical for these types of objects: the time one. This will open a new era for UVOIR studies of INSs, which will play a crucial role in the understanding several open issues in neutron star physics.

\end{document}